\documentclass[12pt]{elsart}
\usepackage{amssymb}
\usepackage{amsmath}
\usepackage{bm}
\usepackage{epsfig}
\usepackage[dvips]{color}
\expandafter\ifx\csname package@font\endcsname\relax\else
 \expandafter\expandafter
 \expandafter\usepackage
 \expandafter\expandafter
 \expandafter{\csname package@font\endcsname}%
\fi

\newcommand{\prd}{Phys. Rev. D}
\newcommand{\aj}{Astron. J. (USA)}

\newcommand{\aap}{Astron. \& Astrophys.}
\newcommand{\physrep}{Phys. Reports}
\newcommand{\prl}{Phys. Rev. Lett.}
\newcommand{\cmda}{Cel. Mech. Dyn. Astron.}
\newcommand{\ijmpd}{Int. J. Mod. Phys. D}
\newcommand{\mnras}{Mon. N. Roy. Soc. London}
\newcommand{\be}{\begin{equation}}
\newcommand{\en}{\end{equation}}
\newcommand{\br}{\begin{eqnarray}}
\newcommand{\er}{\end{eqnarray}}

\renewcommand{\a}{\alpha}
\renewcommand{\b}{\beta}

\begin{document}
\begin{frontmatter}
\title{ Prospects in the orbital and rotational dynamics of the Moon with the advent of sub-centimeter lunar laser ranging}
\author{S.M. Kopeikin\corauthref{cor}}
\address{Dept. of Physics \& Astronomy, Univ. of Missouri,
223 Physics Bldg., Columbia, MO 65211, USA}
\corauth[cor]{Corresponding author}
\ead{kopeikins@missouri.edu}
\author[label1]{E. Pavlis} and \author[label2]{D. Pavlis}
\address[label1]{JCET/UMBC, 523 Research Park Drive, Suite 320, Baltimore, MD 21228, USA}
\address[label2]{SGT Inc., 7701 Greenbelt Rd, Suite 400, Greenbelt, MD 20770-2037, USA}
\author[label3]{V.A. Brumberg\thanksref{footnote1}},
\author[label4]{A. Escapa},
\author[label5]{J. Getino},
\author[label7]{A. Gusev},
\author[label6]{J. M\"uller},
\author[label8]{W.-T. Ni},
\author[label7]{N. Petrova}
\address[label3]{Institute of Applied Astronomy,  10 Kutuzova emb., St. Petersburg, 191187, Russia} 
\address[label4]{University of Alicante, Carr. San Vicente, San Vicente del Raspeig, 03690, Spain}
\address[label5]{University of Valladolid, Prado de la Magdalena, Valladolid, 47005, Spain}
\address[label7]{Kazan State University, 18 Kremlevskaya str., Kazan, 420008, Russia}
\address[label6]{Leibniz University of Hannover, Schneiderberg 50, Hannover, 30167, Germany}
\address[label8]{Purple Mountain Observatory, Chinese Academy of Sciences, 2 Beijing W. Rd., Nanjing, 210008, China}

\thanks[footnote1]{Current address: 100 Norway St., Apt. 5C, Boston, MA 02115-3426, USA}
\begin{abstract}
 Lunar Laser Ranging (LLR) measurements are crucial for advanced exploration of the laws of fundamental gravitational physics and geophysics as well as for future human and robotic missions to the Moon. The corner-cube reflectors (CCR) currently on the Moon require no power and still work perfectly since their installation during the project Apollo era. Current LLR technology allows us to measure distances to the Moon with a precision approaching 1 millimeter. As NASA pursues the vision of taking humans back to the Moon, new, more precise laser ranging applications will be demanded, including continuous tracking from more sites on Earth, placing new CCR arrays on the Moon, and possibly installing other devices such as transponders, etc. for multiple scientific and technical purposes. Since this effort involves humans in space, then in all situations the accuracy, fidelity, and robustness of the measurements, their adequate interpretation, and any products based on them, are of utmost importance. Successful achievement of this goal strongly demands further significant improvement of the theoretical model of the orbital and rotational dynamics of the Earth-Moon system. This model should inevitably be based on the theory of general relativity, fully incorporate the relevant geophysical processes, lunar librations, tides, and should rely upon the most recent standards and recommendations of the IAU for data analysis. This paper discusses methods and problems in developing such a mathematical model. The model will take into account all the classical and relativistic effects in the orbital and rotational motion of the Moon and Earth at the sub-centimeter level. The model is supposed to be implemented as a part of the computer code underlying NASA Goddard's orbital analysis and geophysical parameter estimation package GEODYN and the ephemeris package PMOE 2003 of the Purple Mountain Observatory. The new model will allow us to navigate a spacecraft precisely to a location on the Moon. It will also greatly improve our understanding of the structure of the lunar interior and the nature of the physical interaction at the core-mantle interface layer. The new theory and upcoming millimeter LLR will give us the means to perform one of the most precise fundamental tests of general relativity in the solar system.   
\end{abstract}
\maketitle
\end{frontmatter}
\newpage
%\tableofcontents
\newpage
\section{Introduction}\label{gfree}

After U.S. President George W. Bush announced his "Vision for Space Exploration" in 2004, a plan for new manned lunar missions, NASA elaborated on a program that envisions the construction of a manned lunar base, which will require broad international cooperation \cite{ild}. Success of the human mission to the Moon requires establishing practical navigational network covering the entire surface of the Moon as well as the Earth-Moon interplanetary space. In order to maintain this network as accurate as possible and to prevent its degradation as time passes, it must account for both short and long-term dynamical effects in orbital and rotational dynamics of Earth and Moon whose insufficient knowledge may set limitations on the precision of lunar navigation. Thus, exhaustive studying of rotational and orbital dynamics of the Moon is an essential component of the advanced lunar exploration program from a practical point of view because an accurate lunar ephemeris is critical for future missions to the Moon and the physical librations are critical to landing on the Moon. 

At the same time, precise observation of the lunar rotation gives us invaluable information about internal structure of the Moon, which has a fundamantal importance for understanding of the history of its formation, which in combination with geophysical studies may shed light on the process of formation of the solar system as a whole. Rotational motion of the Moon strongly couples with its orbital motion because of tides leading to capture of the Moon to 1:1 resonance. The tidal response and transition to the resonance depend on the elastic properties of the lunar interior. Therefore, the progress in better understanding of the lunar interior is inseparable from the much more careful exploration of dynamical evolution of the Earth-Moon system.

These questions are being under peer study by a number of national and international lunar and planetary centers and institutes. Today, we have in our disposal intriguing and interesting data on the Moon collected by previous NASA space missions such as Appolo, Clementine \cite{clement}, Lunar Prospector \cite{lupr}, ESA mission Smart-1 \cite{smart}, and the ground-based lunar laser ranging experiment that has been lasting since 1969 \cite{alley1}. Planned space missions Selene \cite{selene}, Chandrayaan-1 \cite{chandr}, Chang'e-1 \cite{chang}, and Lunar Reconnaissance Orbiter \cite{lro} have been proposed and approved to significantly enrich previous information on the Moon due to emerging advanced space technologies, communication facilities and new, sophisticated instrumentation. Physically adequate analysis of data from these satellites requires much better dynamical models of the orbital and rotational behaviour of the Moon as well. 

Lunar Laser Ranging (LLR) is a technique based on a set of corner retro-reflectors located on a visible side of the Moon \cite{alley2} making the natural reference frame for future lunar navigation on the visible (near) side of the Moon. This network can be a crucial component for extrapolation of the lunar positional network on the far side of the Moon with the help of radio-transponders placed on the Moon by robotic lunar expeditions, whose integrity and precision are supposed to be maintained via radio link with a positioning spacecraft (or a constellation of spacecrafts) orbiting the Moon. LLR technique is currently the most effective way to study the interior of the Moon and dynamics of the Moon-Earth system. The most important contributions from LLR include: detection of a molten lunar core and measurement of tidal dissipation in the Moon \cite{wetal};  detection of lunar free librations along with new forced terms from Venus \cite{wdick}; an accurate test of the principle of equivalence for massive bodies (strong equivalence principle also known as the Nordtvedt effect) \cite{muel1}; and setting of a stringent limit on time variability of the universal gravitational constant and (non)existence of long-range fields besides the metric tensor \cite{nord1}. LLR analysis has also given access to more subtle tests of relativity \cite{muel2,muel3,nord2}, measurements of the Moon's tidal acceleration and Earth's precession, and has provided orders-of-magnitude improvements in the accuracy of the lunar ephemeris and three-dimensional rotation \cite{nordik,wtmu,muelwt}. On the geodesy front, LLR contributes to the determination of Earth orientation parameters, such as nutation, precession (including relativistic precession), polar motion, and UT1, i.e. especially to the long-term variation of these effects. LLR contributes to the realization of both the terrestrial and selenocentric reference frames. The realization of a dynamically defined inertial reference frame, in contrast to the kinematically realized frame of VLBI, offers new possibilities for mutual cross-checking and confirmation \cite{muelsk}.

Over the years, LLR has benefited from a number of improvements both in observing technology and data modeling, which led to the current accuracy of post-fit residuals of $\sim$2 cm (see, for example, \cite{apollo} and section 11 in ILRS Annual Report for 2003/2004 \cite{ilrs}).
Recently, sub-centimeter precision in determining range distances between a laser on Earth and a retro-reflector on the Moon has been acheived \cite{apollo2}. As precision of LLR measurements was gradually improving over years from a few meters to few centimeters, enormous progress in understanding evolutionary history of the Earth-Moon orbit and the internal structure of both planets has been achieved. With the precision approaching 1 millimeter and better, accumulation of more accurate LLR data will lead to new, fascinating discoveries in fundamental gravitational theory, geophysics, and physics of lunar interior whose unique interpretation will intimately rely upon our ability to develop a systematic theoretical approach to analyze the sub-centimeter LLR data. This approach should incorporate not only the main Newtonian perturbations but has to treat much more subtle relativitsic phenomena, to decouple effectively orbital and rotational post-Newtonian effects, and to single out annoying spurious solutions of the orbital and/or rotational equations of motion having no physical meaning. The spurious solutions are expected in the description of the lunar motion because the problem is essentially relativistic three-body problem where the third body is the Sun perturbing the lunar orbit significantly. It was proven \cite{bk-ncb,k07,dsx-94} that the relativitsic three-body problem has large freedom in making coordinate gauge transformations from the global barycentric to the local geocentric frame. This gauge freedom brings about ambiguity in the structure of relativistic corrections to the equations of translational and/or rotational motion of the Moon and Earth. Explicit form of the ambiguous terms depends on a specific choice of the gauge conditions imposed on the metric tensor. If a special care is not taken, these gauge-dependent terms can be misinterpreted as physically observable while they just represent a freedom in choosing local coordinates \cite{k07}. This situation is well-known in cosmology where the theory of primordial cosmological perturbations is designed specifically in terms of gauge-invariant quantities avoiding unphysical degrees of freedom \cite{mukh}. The gauge degrees of freedom, existing in the relativistic three-body problem can lead to misinterpretation of gravitational physics of the Earth-Moon system, thus, degrading value of extremaly accurate LLR measurements for unambiguous planning of lunar exploration and establishing a selenodesic navigational network. 

The primary objective of the improved theory of the lunar motion has to be developing a new analytic approach to the numerical computer model of the orbital motion and intrinsic rotation of the Moon and Earth that will be gauge-invariant and include modelling of all classical (lunar interior, Earth geophysics, tides, asteroids, etc.) and relativistic effects having amplitude equal to or slightly less than 1 millimeter. The gauge freedom in the three-body problem (Earth-Moon-Sun) should be carefully examined by making use of the scalar-tensor theory of gravity and the principles of analytic theory of relativistic reference frames in the solar system \cite{bk-rf,dsx-1991,kv-pr} that was adopted by the XXIV-th General Assembly of the International Astronomical Union \cite{iau2000} as a standard for data processing of high-precision astronomical observations. The analytic approach will help to exclude the spurious gauge-dependent modes from the orbital and rotational motion of the celestial bodies which will allow us to adequately interpret new observational data from LLR and lunar orbiters, thus, making our knowledge of various physical aspects of the Earth-Moon system as well as the lunar interior, robust and unambiguous. 

It should be clearly understood that any coordinate system can be used for data processing since any viable theory of gravity obeys the general principle of relativity. For this reason, we are not saying that there is a privileged coordinate frame for rendering analysis of the LLR data with sub-centimeter accuracy. However, there exist effects which are directly associated to the relativistic space-time transformations between various post-Newtonian frames. These effects have pure coordinate origin, can not be counted as physical, and as such, they must be excluded from the number of measured phenomena.

Existing theories of the lunar motion are either numerical \cite{le,pitj} or semi-analytical \cite{chapron}. They consist of three major ingredients: 
\begin{itemize}
\item[(1)] Einstein-Infeld-Hoffmann (EIH) equations of motion of the Moon, Earth, Sun, and other planets of the solar system; 
\item[(2)] rotational equations of motion; 
\item[(3)] equations of propagation of light rays.
\end{itemize}
 Orbital equations are written down and numerically integrated in the barycentric reference frame of the solar system. This barycentric approach does not control the contribution of the gauge-dependent modes existing in the solution of the equations of motion of the Moon with respect to Earth \cite{bk-ncb,dsx-94,k07}. Because the gauge-dependent modes do not carry physically relevant information, it remains incomprehensible if results of fitting of the LLR data to the lunar numerical ephemerides in the barycentirc frame provide us with physically-consistent approach to deeper exploration of the lunar interior and establishing the selenodetic navigational network. The hidden dependence of the barycentric approach on the gauge modes brings uncertainty in our understanding to what extent Einstein's theory of general relativity was tested with LLR. Unfortunately, the previous research papers devoted to discussion of LLR tests of relativity \cite{nord1,nord2,nord3} did not separate to full extent the ingenious post-Newtonian perturbations from ordinary coordinate effects of the post-Newtonian three-body problem which are unobservable artefacts of gravitational physics of the Earth-Moon system. New advanced approach to the construction of theory of the lunar orbital and rotational motion on the basis of the IAU2000 resolutions \cite{iau2000} will help us to resolve this issue and focus on the detection of covariant physical effects in orbital/rotational motion of the Moon existing independently of the choice of coordinates. 

Practical purpose of the new lunar theory is
to develop a computer-based lunar ephemeris program which will be able to keep precision in determination of positions and velocities of spacecrafts up to 1 millimeter and 1 millimeter/second respectively on long time intervals, presumably in several decades. In order to acheive this goal one has to incorporate to the practical model of the orbital/rotational motion of the Moon, the analytic solution of equations describing propagation of light from Earth to the Moon and back. This solution establishes a basis for optical/radio communication link between station on the Earth and a communication device in space and/or on the Moon, and consists of two parts describing respectively the coordinate distance between the emitter of light (laser) and a retroreflector on the Moon, and the static relativistic correction known as the Shapiro time delay \cite{shapiro}. The light-ray equation is a necessary ingredient in developing the covariant analysis of the LLR data. Our advanced approach to the entire theory of the lunar motion makes it necessary to extend the previously existed analysis of the light-ray propagation in order to include velocities of the laser and the retroreflector to the Shapiro time delay. Additional motivation for improving the theory of light-ray propagation in the gravitational field of the solar system is an exciting possibility to use results of the differential VLBI, which will be obtained in next few years from observations of Japanese \cite{japan} and Chinese \cite{china} lunar orbiters, for comparison with the results of LLR data analysis. Since precision of the lunar differential VLBI is expected to be comparable with that of LLR technique \cite{vlbi-llr} or may be even better \cite{lunar-vlbi}, this comparision will allow us to set stringent limitations on parameters of the multi-layer model of the lunar interior and on Moon's orbital/rotational motion making navigation of regular space flights to the Moon much more precise.

\section{Orbital Dynamics}
The advanced post-Newtonian theory of the orbital dynamics of the Earth-Moon system should include the following elements: 
\begin{enumerate}
\item construction of a set of astronomical frames decoupling description of the orbital dynamics of the Earth-Moon system from that of the rotational motion of Earth and Moon with full account for the post-Newtonian corrections to the Newtonian theory and elimination of the gauge modes;
\item relativistic definition of the integral parameters like mass, the center of mass, the quadrupole moment, oblateness, etc. of Earth, Moon, and other solar-system bodies that excludes terms having pure kinematic origin due to inadequate choice of the origin of the local coordinates;
\item derivation of relativistic equations of motion of the center-of-mass of the Earth-Moon system with respect to the barycentric reference frame of the solar system taking into account gravitational interaction of the Earth-Moon system intrinsic multipoles with tidal gravitational field of the Sun and other planets;
\item derivation of relativistic equations of motion of Earth (Moon) with respect to the barycentric reference frame of the Earth-Moon system taking into account gravitational interaction of Earth (Moon) intrinsic multipoles with tidal gravitational field of the Moon (Earth), Sun and other planets;
\item derivation of relativistic equations of motion of a lunar orbiter with respect to the selenocentric reference frame having an origin at the center of mass of the Moon;
\item calculation of partial derivatives with respect to parameters of the orbital theory for implementation in the computer code fitting sub-centimeter LLR data to the theory;
\item comparision of the lunar part of the numerical code with currently existing JPL software used for LLR data processing;
\item fitting LLR data with the new model to obtain physical parameters of our theoretical model of the orbital motion of the Moon, and to test predictions of general relativity at new, challenging experimental level.  
\end{enumerate}

Relativistic theory of gravity is required to accomplish these steps in order to make the orbital theory of the lunar motion compatible with the sub-centimeter accuracy of the new LLR observations. The most suitable framework is a scalar-tensor theory of gravity with two parameters -- $\beta$ and $\gamma$, which are the most representative parameters of the parametrized post-Newtonian (PPN) formalism \cite{willbook}. These two parameters describe presumable deviations from general theory of relativity due to the presence of a hypothetical long-range scalar field, and are also convenient tool in testing correctness of the computer coding of the theoretical model. Equations of the scalar-tensor theory have been used at Jet Propulsion Laboratory (JPL) \cite{standish1,standish2,standish3,standish4,standish5}, and other space research centers \cite{cfa,pitjeva,sitarski} for constructing the barycentric ephemerides of the solar system bodies (including the Moon) giving their positions and velocities with respect to the barycentric reference frame of the solar system having origin at the center of mass of the solar system. 

Gravitational field in the scalar-tensor theory of gravity is described by the metric tensor $g_{\a\b}$ and a long-range scalar field $\phi$ loosely coupled with gravity by means
of a function $\theta(\phi)$.
Field equations in the
scalar-tensor theory are derived from the action \cite{willbook}
\be
     S=\frac{c^3}{16\pi}\int\biggl(\phi R                  \label{10.1}
     - \theta(\phi)\frac{\phi^{,\alpha}\phi_{,\alpha}}
     {\phi} - \frac{16\pi}{ c^4} {\mathcal L}(g_{\mu\nu},\,\Psi)\biggr)\sqrt{-g}\; d^4 x\; ,
\en where the first, second and third terms in the right side of equation (\ref{10.1}) are
the Lagrangian densities of gravitational field, scalar field and matter
respectively, $g={\rm det}[g_{\alpha\beta}]<0$ is the determinant of the metric tensor
$g_{\alpha\beta}$, $R$ is the Ricci scalar, $\Psi$ indicates dependence of the matter Lagrangian ${\mathcal L}$ on
matter fields, and $\theta(\phi)$ is
the coupling function describing "kinematic mass" of the scalar field, which is kept arbitrary. This makes the theory covering a wide range of alternative scalar-tensor theories of gravity that are
sufficiently universal.
For the sake
of simplicity one postulates that the intrinsic
potential of the scalar field is identically zero so that it does not interact
with itself.
This is because this potential does not lead to
measurable relativistic effects within the boundaries of
the solar system \cite{willbook}. On the other hand, the non-linearity of the scalar field (and, hence, function $\theta(\phi)$) can be important in strong gravitational fields of neutron (or quark) stars, and its
inclusion to the theory
leads to interesting physical consequences described, for example, in \cite{1992CQGra...9.2093D,1993PhRvL..70.2220D,1993PhRvL..70.2217D} which can be likely tested at the new, excited level of accuracy of LLR measurements besides binary pulsars.

Physically adequate relativistic description of the lunar motion is not conceivable without a self-consistent theory of relativistic reference frames in the solar system. The solar system has a hierarchic structure associated with the hierarchy of masses of the solar system bodies \cite{jpldata}. The most massive body of the solar system is the Sun. Planets are orbiting the Sun because their masses are significantly smaller so that they do not gravitationally attract the Sun strong enough and it stays all the time very close to the center of mass (barycenter) of the solar system. Nevertheless, major planets of the solar system, like Jupiter and Saturn, attract the Sun fairly significant to render its revolution around common solar-system barycenter (SSB) that is not located at the Sun's center of mass \cite{jpldata}. A global, solar-system barycentric frame is required to describe the orbital motion of Sun and planets around the SSB. On the other hand, rotational motion of the Sun and planets is more natural to describe in a local frame associated with each of the bodies. Many planets have their own satellites orbiting the planet. A planet with all its satellites comprise a sub-system of the solar system that can be considered as being gravitationally-isolated from the rest of the solar system because the principle of equivalence tells us that the local frame is almost inertial up to small perturbations associated with the tidal gravitational field of external masses. Depending on the ratio of masses of the planet and its satellites, it is convenient to introduce a local coordinate frame associated with the barycenter of the sub-system made of the planet and its satellites. This is because motion of the barycenter of the sub-system around the SSB is very close to the Keplerian ellipse while the planet and its satellites move periodically (oscillate) around the barycenter of the sub-system. In other words, the hiearchic structure of the coordinate frames in the solar system brings about mathematical decomposition of the orbital motion of bodies in the solar system in a set of fundamental Fourier harmonics whose amplitudes and frequencies one has to determine from observations. 

From the hierarchic point of view the theory of the orbital motion of Moon should include construction of three local coordinate frames attached to: (1) the Earth's center of mass (geocenter), (2) to the Moon's center of mass (selenocenter), and (3) to the barycenter of the Earth-Moon sub-system. It is worth emphasizing that the Earth and Moon can not be considered as point-like, structureless bodies, and their orbital motion must be disentangled from their intrinsic rotations. Therefore, the geocentric frame is required to describe rotational motion of Earth and motion of artificial satellites orbiting Earth. The selenocentric frame is introduced to describe rotational motion (precession, nutation, libration, polar wobble, etc.) of the Moon and the orbital motion of spacecrafts around the Moon. The local frame with the origin at the Earth-Moon barycenter (EMB) is introduced to describe the relative motion of Moon with respect to Earth. The global SSB frame is used to describe the orbital motion of the EMB frame with respect to the solar-system barycenter, and it includes all tidal effects of the Sun and other planets. 

Relativistic theory of gravity brings about additional justifications in favor of such a hierarchic fragmentation. It naturally arises on space-time manifold because of its differential structure described in terms of a set of local coordinates and the geometric objects such as the metric tensor, the Christoffel connection, and the curvature (Riemann) tensor \cite{mtw}. Let us consider a group of bodies which are gravitationally bounded to each other and form a sub-system of an N-body system that interacts with the rest of the system only through tidal gravitational field. The N-body system is comprised of a number of the sub-systems. It is natural to introduce a single global frame covering the whole N-body system, and a set of local inertial coordinates associated with each of the sub-systems. In the first approximation the orbital motion of bodies in each of the sub-system is described by a linear combination of a solution of the ordinary differential equation of time-like geodesic of the body's center of mass, and that of the equation of the geodesic's deviation associated with the tidal forces from other sub-systems of bodies that are external to the sub-system under consideration. We shall focus in this proposal on the sub-system consisting of Earth and Moon. External sub-systems are formed by other planets and the Sun which affect the orbital motion of the Moon with respect to Earth only by means of tidal forces. This consideration is valid not only in the Newtonian but in the post-Newtonian (relativistic) approximation as well \cite{bk-rf,brum-book,kop-rf}, if one is careful enough in elimination of the spurious gauge-dependent terms from the solutions of the orbital equations of motion.

It is clear that the relativistic equation of orbital motion of the Moon around Earth can be written in arbitrary coordinates (for example, in the barycentric coordinates of the global frame) but it has the most simple mathematical form in the local coordinate frame (because the Moon orbits the Earth) where each term of the equations has transparent physical meaning in an appropriately chosen gauge  \cite{bk-ncb,dsx-94}. Physically adequate construction of the set of local frames connected to a global frame by a differentiable space-time coordinate transformation helps us to single out and to eliminate a great deal of spurious, gauge-dependent relativistic effects which are artefacts of the residual gauge freedom of the gravity field equations. It is well-known that the spurious, gauge-dependent radial oscillations of the Earth-Moon coordinate distance inhibited earlier developments of the relativistic theories of the lunar motion \cite{vab,baier}. Their amplitude amounts to a few meters and a number of well-established researchers really believed in their physical existence and measurability \cite{baier,1982A&A...116...75L,1982hper.coll..217L}. Though, the spurious character of some of these radial-oscillation terms was well-understood \cite{1973PhRvD...7.2347N,1981rcse.conf..283B,sof}, there are still many other non-physical terms which are explicitly present in the coordinate description of lunar motion when it is given either with respect to the global frame of the solar system or with respect to the local coordinate frame which is constructed improperly (see, for instance, an interesting exchange of opinions on this subject in \cite{k07,nord3,quq}).

Our post-Newtonian theoretical approach to the description of the lunar motion divides motion of Earth and Moon in several major components. First, we separate the motion into the orbital motion of the center of mass of the Earth-Moon system around the solar-system barycenter (SSB), and the orbital motion of the Moon and Earth with respect to the local frame associated with the Earth-Moon barycenter (EMB). This separation was a key ingredient of the classical, Hill-Brown theory of the lunar motion \cite{hill,brown1,brown2}, and it is important to preserve this separation in relativistic terms of the scalar-tensor theory of gravity as demonstrated in \cite{kv-pr}. The problem with the barycentric approach to the construction of the lunar theory adopted by JPL in the construction of DE/LE series of ephemerides of planets, the Sun, and the Moon is that it uses a single, global coordinate frame for numerical integration of the equations of motion of bodies. Numerical integration is highly accurate as concerned the coordinate description of the orbital motion. But the problem is that in relativity and geodesy the coordinates are not directly observable quantities. They enter solution of the relativistic equation of propagation of light rays which are used for practical LLR (and other astronomical) observations. Numerically integrated equations of motion of the Moon have a great number of spurious oscillatory modes in the barycentric coordinate distance between Earth and Moon. In a perfect (analytic) scenario the spurious terms in the equations of motion of the Moon must mutually cancel out with corresponding spurious terms present in the solution of the equations of light-ray propagation. JPL integrator code was not designed to control the degree of the cancellation of the pure coordinate degrees of freedom. Hence, it is conceivable to expect that the current numerical (LE) ephemerides of the Moon may have some parasitic degrees of freedom corresponding to the relativistic gauge transformations which generate fictitious effects in description of the orbital motion of the Moon that are really absent in the ranging data \cite{k07}. It is important to pin down (label) these parasitic effects in order to eliminate them from physically observable parameters. Some research in this direction has been conducted by Brumberg and Simon \cite{brsim}. The analytic theory of the reference frames and the equations of motion \cite{bk-ncb,bk-rf,kv-pr} should be extended in such a way that will unambiguously parametrize the gauge terms of the relativistic three-body problem, thus, helping us to re-establish control over all spurious gauge-dependent modes in the numerically-integrated ephemerides of the lunar orbital motion. This will allow us to uniquely interpret LLR observations taken with sub-centimeter accuracy avoiding undesirable blunders in identification of real physical effects in the orbital motion of the Moon which can jeopardize precise orbital and landing navigation of the lunar spaceships. 

\section{Rotational Dynamics}

Relativistic theory of the reference frames \cite{bk-ncb,bk-rf,dsx-1991,kv-pr} which we suggest for doing the sub-centimeter LLR data processing, maximally separates physical effects of intrinsic rotation of the Moon and Earth from their orbital motions not only in the Newtonian approximation but also taking into account the post-Newtonian corrections. This is important to ensure that unphysical modes associated with the freedom of choosing the coordinate systems do not mix with physical libration and nutation. Picking up a dynamically non-rotating local coordinate frame \cite{bk-rf} associated with the Moon (Earth) one can cast the post-Newtonian rotational equations of motion for the body's spin $S^i$ in the following form (see section 10 of the paper \cite{kv-pr} and \cite{dsx-3})
\begin{equation}
\label{2}
\frac{d{\mathcal S}^i}{du}=T^i+c^{-2}\Delta T^i\;,
\end{equation}
where $c$ is the fundamental speed of the Minkowski space-time,
\begin{equation}
\label{3}
T^i=\epsilon_{ijk}\sum_{n=0}^{\infty}\left[{\mathcal I}^{ja_1a_2...a_n}\left(Q_{ka_1a_2...a_n}-c^{-2}\dot Z_{ka_1a_2...a_n}\right)+c^{-2}{\mathcal S}^{ja_1a_2...a_n}C_{ka_1a_2...a_n}\right]\;
\end{equation}
is the gravitational torque consisting of the Newtonian terms (proportional to $Q_{ka_1a_2...a_n}$) and the post-Newtonian perturbations (proportional to $C_{ka_1a_2...a_n}$), and $\Delta T^i$ is the torque caused by the presence of the scalar field that is absent in general relativity. Here ${\mathcal I}^{a_1a_2...a_n}$ and ${\mathcal S}^{a_1a_2...a_n}$ are the mass-density and current-density multipole moments of the rotating body, whereas $Q_{a_1a_2...a_n}$ and $C_{a_1a_2...a_n}$ are external multipole moments of the metric gravitational field associated with its gravitoelectric and gravitomagnetic properties respectively \cite{dsx-1991,kv-pr}.

The local coordinate frame associated with the Moon (Earth) is dynamically non-rotating, has its origin at the center of mass of the Moon (Earth) defined at the post-Newtonian precision.  Gravitational torque, $T^i$, of external masses (Sun, planets) reveals itself only in the form of the tidal multipole terms, while the gravitational torque $\Delta T^i$ caused by the scalar field is proprtional to the differences $\gamma-1$ and $\beta-1$ multiplied with the orbital acceleration of the Moon (Earth) and vanishes in general relativity. 

Spin ${\mathcal S}^i$ is defined taking into account the post-Newtonian corrections  \cite{dsx-1991,kv-pr}. Nevertheless, it can be represented (similar to the Newtonian theory) as a product of the relativistic moment of inertia and the angular velocity of rotation plus small relativistic corrections whose exact structure is still a matter of discussion \cite{tg,pnomega}. Though these relativistic corrections are not important at the level of LLR measurements in few centimeters, they may become important when one reach 1 millimeter in the accuracy of ranging measurements to the Moon. 

Equation (\ref{3}) clearly demonstrates that 
\begin{itemize}
\item[(1)] the free rotational motion (free precession, polar wobble, nutation, libration) is governed basically by the distribution of internal forces and matter density inside the body under consideration; 
\item[(2)] the impact of the orbital motion on rotation is important only to that extent to which the tidal forces are not ignored;
\item[(3)] the spin-orbit interaction is realized through the coupling of the external tidal moments $Q_{a_1a_2...a_n}$ and $C_{a_1a_2...a_n}$ with the internal multipole moments ${\mathcal I}^{ja_1a_2...a_n}$ and ${\mathcal S}^{a_1a_2...a_n}$ respectively, which "tensorial" orientation in space (for example, amplitude and orientation of tidal buldge) and time behaviour depend on the internal structure of the rotating body characterized by Love's numbers;
\item[(4)] the spin-orbit interaction is affected by kinematic effects like the orbital acceleration or velocity of the rotating body neither in the Newtonian nor in the post-Newtonian approximations.    
\end{itemize} 

This well-defined separation of the post-Newtonian coordinate effects from the physical effects in orbital and rotational equations of motion is an essential element of our theoretical approach \cite{bk-rf,kv-pr}. Earlier developments of the post-Newtonian theory of rotational motion of celestail bodies suffered from incomplete separation of orbital and rotational dynamics at the level of the post-Newtonian corrections (see, for example, \cite{br-book,capor}) so that some gauge degrees of freedom in the definition of the post-Newtonian spin and/or torque might be mistakenly interpreted as physical. This important issue is known in relativistic celestail mechanics as "the effacing principle" \cite{damour83} and its exhaustive discussion is given in \cite{kv-pr,damour87,kv-mg11}. Validity of the effacing principle was proven in general relativity up to the post-Newtonian corrections of the order of $(v/c)^{10}$ \cite{brud,damour83,kop85,kg86} but in the scalar-tensor theory of gravity it is valid only up to the terms of the order of $(v/c)^{6}$ \cite{kv-pr,kv-mg11}. The effacing principle should not be confused with the strong principle of equivalence which is about equality between the inertial and gravitational masses of celestial bodies \cite{dicke}.

Currently, it is not quite clear whether the Newtonian theory is sufficient for complete interpretation of the rotational data of the Moon and Earth, or the post-Newtonian corrections should be earnestly taken into account. The fact that relativistic corrections might be important in the rotational theory of the Moon and Earth, has been recently pointed out in several papers \cite{bbf,vok,vokr,brg,ksr1}. For example, apart from the effects common to the two reference systems (including the well-known geodetic precession of the lunar orbit \cite{wnd}), the lunar reference frame undergoes an additional precession of 28.9 milli-arcsec/century \cite{vok}. This value is theoretically within the range of the lunar laser ranging (LLR) technique attaining precision of 1 millimeter, but it should be decorrelated from other secular effects of selenophysical origin. Existence of the rotational-orbital 1:1 resonance in the lunar dynamics imposes a specific constrain on the de Sitter and Lense-Thirring relativistic nutations making them almost equal to each other and hard to observe \cite{vokr}. M\"uller \cite{muelphd} made a significant effort to incorporate relativistic effects to the rotational-orbital dynamics of the Moon by making use of Thorne-Hartle \cite{th} and Brumberg-Kopeikin \cite{bk-ncb,bk-rf} formalism and tested it in his LLR software. He could show that their impact was fairly small and removed those terms again from his LLR software, because their computation was quite time consuming. However, the sub-centimeter LLR demands to reconsider this problem at new theoretical level. Specific and systematic work in this direction has been conducting by research groups in Germany and China \cite{xws1,xwsk,xws2} and it would be interesting to study this problem in more detail in application to the real LLR data.

An attempt to construct accurate post-Newtonian theory of the Moon's spin-orbital motion by numerical integration has been undertaken in \cite{boisv}. Several phenomena capable of producing effects of at least $10^{-4}$ arcseconds in the lunar physical librations have been included and analysed, in particular: (i) de Sitter precession of the Earth reference frame, and (ii) quasi-Newtonian torques acting on the lunar physical librations. The global behaviour of the librations shows that their resulting magnitude is, indeed, of the order of a few $10^{-4}$ arcseconds and may reach one milli-arcsecond. This works reveals that unlike the very faint relativistic contribution to the Earth's rotation, the corresponding relativistic terms in the Moon's rotational motion are not negligible even with respect to the present observational accuracy of the lunar laser data. Consequently it is appropriate to consider them in models adjusted to the observations. If results of these preliminary studies are true (and we are going to check them), the sub-centimeter LLR will be capable to detect a number of new, interesting effects leading to better understanding of the fundamental issues in the definition of spin and rotation in general relativity. 

\section{Lunar Librations}
Besides challenging relativistic aspects, the advanced theory of the lunar rotation should elaborate moree deeply on the following classical topics of the lunar science: 
\begin{enumerate}
\item development of the analytical theory of rotation of two/three-layer Moon and prediction of lunar physical libration;
\item estimation of the internal energy dissipation and modeling the long-period mechanism maintaining the free libration;
\item analysis of the evolution of the core-mantle boundary with possible reconstruction of gravitational and viscous-mechanical interaction between the core and the mantle; 
\item calculation amplitudes and frequencies of the free and forced nutation of the lunar liquid core;
\item derivation of the improved values of the Love numbers and the tidal quality-factor Q for the Moon.
\end{enumerate}

The internal structure of the Moon is one of the most puzzling among the terrestrial planets. The space upcoming missions as well as ground-based LLR measurements will play an important role in constraining our understanding of the structure, formation, and evolution of the Moon. The development of a complete theory of the spin-orbit motion of the Moon is an essential complement to observational data and will improve significantly our knowledge of the planet. Prior work concerning the effect of core-mantle couplings on the rotation of the Moon has assumed that the obliquity of the Moon is equal to zero and that its orbit is Keplerian. Future work has to deal with the core-mantle interaction in a more realistic model of the orbital and rotational motions of the Moon strating from the existing JPL model of the solar system. We shall study the dynamical behavior of the rotational motion of the Moon considered as consisting of a solid mantle including a liquid and/or a solid core. The liquid core and the mantle are assumed to be coupled through an inertial torque on the ellipsoidal core-mantle boundary. We are going to determine Moon's rotation for a large set of parameters characterizing various models of the interior structure of the Moon to be able to identify and to clarify the impact of the core motion on the lunar librations. We shall present a comparative study of the librations resulting from different models of the internal structure. 

Previous research on lunar librations showed that they have sensitivity to interior structure, physical properties of the mantle, and energy dissipation. The second-degree lunar Love number $k_2$ has been detected with an accuracy of 11\% \cite{wbr}. Lunar tidal dissipation is strong, and its Q-factor has a weak dependence on tidal frequency \cite{efr3}. A liquid core of about 20\% the Moon's radius is indicated by the dissipation data \cite{wit}. Evidence for the oblateness of the lunar fluid-core/solid-mantle boundary is getting stronger \cite{wbr1} and the advent of sub-centimeter LLR can give a robust confirmation of its existence.  

Our analytic theory of rotational motion of a deformable model of the Moon will be constructed on the basis of the Hamiltonian form of equations by making use of a stratified two-layer model (elastic mantle + fluid core) and the Andoyer variables \cite{kinosh1,kinosh2,efr1,efr2}. This approach is physically meaningful, has been practically applied for the model of elastic Earth with a liquid core \cite{getino,fer}, and has also been extended to the study of the rotation of a three-layer Earth model \cite{escapa}. The Andoyer variables are capable to describe precession and nutation of the elastically-deformable Moon along with the dissipative effects caused by tides and core-mantle interaction which are important for correct evaluation of the overall energy loss in Moon's interior. However, since they are not osculating elements of the perturbed equations of rotational motion (because the Hamiltonian depends on velocities), a special care should be taken in order to segregate physical modes of librations from the gauge oscillations which are generated by the canonical transformations in the Hamiltonian theory applied improperly \cite{efr2,efr4}.

The asymptotic methods of the Hamiltonian theory are planned to be used for obtaining analytic solution of the rotational equations of motion of the Moon in the form of trigonometric series in the spirit of works \cite{br-ant,ivan} advocating symbolic mathematical calculations. The important advantage of the symbolic form of the solution is the explicit analytic dependence of the librational variables not only on time, but also on the parameters of the lunar model (liquid core's radius and ellipticity, viscosity at the core-mantle boundary, elasticity, etc.) which can be determined from LLR observations by making use of a comparative analysis \cite{chapr}. Another advantage of the analytic approach is that it perfectly matches with the mathematical techniques that are going to be implemented in the data analysis of the lunar space missions Selene \cite{selene,j1,j2,j3} and Chang'e-1 \cite{chang,e-1} whose simulation analysis orbit determination and lunar gravity recovery is presently carried out by making use of GEODYN computer code \cite{geodyn}.  

Three free libration modes are of a particular interest to confirm and/or to study with much better precision with the sub-centimeter LLR: (1) the Chandler wobble (CW), (2) the free precession (FP), and (3) the free core nutation (FCN). 

The Chandler wobble is a motion of the rotation axis of the Moon around its dynamical figure axis due to the bulges of the lunar body. It is the only global rotational mode for completely solid planet. For the Moon it has a long period 74.63 year in a frame tied to the Moon and is prograde (i.e. in the direction of lunar rotation). This mode was detected from LLR observation as 3" to"8" elliptical component in the oscillation \cite{nwi} and it may be related to effects at the core/mantle interface that can be understood with the aid of Yoder \cite{yoder1,yoder2} turbulent boundary layer theory.

The free precession corresponds to the prograde circular motion of the actual pole  about its mean position with radius 0".022 (about 18 cm at the lunar surface) and the period of 24.16 years. In a space-fixed frame the motion is retrograde with a period of 80.77 years \cite{nwi}. 

The free core nutation represents a differential rotation of the liquid core relative to the mantle \cite{fer,fcn1,fcn2,fcn3}. This mode can exist only if a core is liquid. In this case it should have a quasi-diurnal period in the frame connected to the Moon and be retrograde. The lunar FCN would have a long period in space of about 144 year, if dynamical figure of the core is similar to that of the mantle \cite{pg} or about 186 year for the axially symmetric core with ellipticity $\sim 4\times 10^{-4}$ \cite{bgp}. Values of these frequencies have been computed by means of solving corresponding equations of rotation for a two-layer Moon. They depend on the size of the lunar core, its chemical composition, and dynamic flattening. Hence, finding the lunar FCN with LLR technique will allow us to get precise numerical estimates of these essential parameters of the lunar interior.

\section{Earth-Moon Formation and Evolution}

Most scenarios for the formation of the Moon place the Moon near Earth in low-eccentricity orbit in the equatorial plane of Earth \cite{tw}. GEODYN code can examine the dynamical evolution of the Earth-Moon system from such initial configurations during the early evolution of the system to find out if strong orbital resonances are encountered. Passage through these resonances can excite large lunar orbital eccentricity and modify the inclination of the Moon to the equator. A period of large lunar eccentricity would result in substantial tidal heating in the early Moon, providing a heat source for the lunar magma ocean. The resonances may also play a role in the formation of the Moon.

Following the ideas put forward in \cite{kagan-maslova} it is reasonable to study a stochastic model of the tidal evolution of the Earth-Moon system based on two assumptions: (1) tidal energy dissipation occurs in the ocean, and (2) a change of tidal energy dissipation in the past is connected with an alteration of the ocean spectrum due to the continental drift. This may help to understand in more detail the time-scale problem for the tidal evolution of the Earth-Moon system rised by Goldreich \cite{gold1,gold2}. A special attention should be given to the possible correlation between existing estimates of the duration of the tidal evolution of the Earth-Moon system and ocean tidal energy dissipation.

The resonances arising in the Earth-Moon system at the different stages of its dynamic history should be investigated both numerically (GEODYN, PMOE 2003) and analytically using the modern mathematical theory of the resonant phenomena. Conditions of the resonance capture are not quite clear and should be explored along with the analysis of the subsequent evolution of resonant movements. Particular attention should be given to the analysis of formation of synchronous rotation of the Moon - the resonance 1:1 - between the orbital and rotational movements of the early Moon.

Consequences of the prolonged stay of the Earth-Moon system in 1:1 resonance onto the processes in the lunar interior have not been studied to full extent so far and have to be explored. In particular, the rate and amount of energy generated due to non-stationary tidal deformations in resonant movements should be evaluated. Correct estimate of the generated energy is important for analysis and/or explanation of such phenomena as:
(1) dynamo-mechanism of generation/dissipation of magnetic field of the Moon in the past and present epoch, (2) differential rotation between core and mantle of the Moon bringing about the effect of the lunar free core nutation (FCN), (3) convective motion of the lunar interior initiated by the Moon's core-mantle friction leading to ascending plume intrusions that can serve as  possible sources of gravitational anomalies on the far-side of the Moon.

The best way to study the Earth-Moon system dynamical history is by numerical simulations with more realistic rates of tidal evolution estimating the lunar interior heating due to the tidal deformations in the resonance motions. Models of early spin - orbital evolution of the Earth-Moon system should take into account the capture in resonant rotation and the differential rotation of a lunar core. Examples of these resonances are the "evection" and "eviction" discovered in \cite{tw}. Additional resonances can emerge if one takes into account heating of magma ocean and formation of convection structures of a lunar mantle where an extreme transfer of thermal energy and of substance in the form of ascending plume-intrusion may take place. Enhanced lunar dissipation causes the lunar eccentricity to decay but may also result in a regression of the semimajor axis, allowing a new encounter with the resonance. For this encounter, capture is almost certain, provided that the rate of tidal evolution is small enough. After capture, evolution in the resonance excites the mutual inclination to values as large as $12^o$ \cite{tw}. Subsequent evolution of the system can take the system to the present state of the Earth-Moon system. This scenario can be effectively tested with GEODYN code thus providing a possible resolution of the mutual inclination problem within conventional tidal models.

\section{Conclusions}
The National Aeronautics and Space Agency is examining several approaches to meet safety and navigational requirements for spacecraft in lunar orbit, in transit to or from the Moon, and for personnel on the lunar surface.
The proposed work is supposed to significantly contribute to the solution of this practical problem:
\begin{itemize}
\item extended GEODYN model of the reference frames and that of the orbital motion of the Moon around Earth will allow us to navigate spacecrafts at any point between Earth and Moon and in the vicinity of the Moon with unsurpassable accuracy, thus, increasing safety of human missions to the Moon;
\item the improved Hamiltonian theory of the lunar rotation and librations combined with the GEODYN-processed sub-centimeter LLR data will allow us to improve the accuracy of selenodesic network on the near side of the Moon up to a few meters by more precise matching of the lunar photographic observations with positions of the cube-corner reflectors on the Moon \cite{alesh}. This will later facilitate extrapolation of the selenodesic network on the far side of the Moon;
\item advanced investigation of the lunar interior structure with sub-centimeter LLR will significantly improve our understanding of the lunar tectonics and frequency of occurrence of possible seismological events (moon-quakes) generated by the dissipative processes of energy interaction between the fluid core and mantle.      
\end{itemize}

More sophisticated theoretical approach to the problem of orbital and rotational motions of the Moon, described in previous sections, is promising for making significant progress in advanced exploration of the lunar interior. Sub-centimeter LLR data processed with the enhanced GEODYN numerical code would allow us:  
\begin{itemize}
\item  to obtain further, more convincing evidences of existence of the lunar fluid core and determine its radius;
\item to derive precise numerical values of Love numbers and other parameters (moduli) characterizing elastic properties of the Moon;
\item to solve the problem of the initial conditions for the coupled spin-orbit equations describing temporal evolution of the lunar orbit as it passes through resonances;
\item to evaluate the amplitude and frequencies of various harmonics in the rotational motion of the multi-layer Moon: the Chandler wobble, the mantle free precession and nutation, and the free core nutations which depend on the distribution of density, ellipticity, chemical composition, viscosity, and aggregative state of the core, so that these characteristics can be determined from LLR observations;      
\item to analyse dissipative processes and convective turbulence on the boundary layer between the core and mantle, caused by differential rotation of the fluid core interior;
\item to calculate characteristic times of dissipation of the free librational modes for different values of density, viscosity, and other characteristics of the core.  
\item to advance understanding of the lunar topography and to explain the origin of gravitational anomalies (mascons) at later stages of selenological evolution of the Moon.
\end{itemize}

Sub-centimeter LLR and new GEODYN code will be able to significantly improve the test of general relativity in the Earth-Moon system by setting stronger limitations on parameters $\gamma$ and $\beta$ of the PPN formalism. Detection of new relativistic effect are highly plausible. Among them is the gravitomagnetic precession of the Moon and/or its orbit with respect to ICRF, secular and periodic effects caused by relativistic quadrupole moment of the Earth, tidal gravitomagnetic (periodic) effects, presumable violation of the strong principle of equivalence. Sub-centimeter LLR will be also able to lower the limit on the density of stochastic gravitational-wave background in the frequency range $10^{-5}\div 10^{-7}$ Hz.

\section{References and Citations}

\end{document}